\documentclass[twocolumn,fp]{jpsj3}

\usepackage{txfonts}
\usepackage{color}
\usepackage{amsmath}
\usepackage{mathrsfs}
\usepackage{amssymb}
\usepackage{amsfonts}
\usepackage{braket}
\usepackage{widetext}
\usepackage{graphicx}
\usepackage{ulem}

\begin{document}
\title{Quantum Hall Effect of Massless Dirac Fermions and Free Fermions \\in Hofstadter's Butterfly}

\author{Nobuyuki Yoshioka, Hiroyasu Matsuura, and Masao Ogata}
\inst{Department of Physics, University of Tokyo, 7-3-1 Hongo, Bunkyo-ku, Tokyo 113-0033, Japan}

\abst{We propose a new physical interpretation of the Diophantine equation of $\sigma_{xy}$ for the Hofstadter problem. First, we divide the energy spectrum, or Hofstadter's butterfly, into smaller self-similar areas called ``subcells'', which were first introduced by Hofstadter to describe the recursive structure. We find that in the energy gaps between subcells, there are two ways to account for the quantization rule of $\sigma_{xy}$, that are consistent with the Diophantine equation: Landau quantization of (i) massless Dirac fermions or (ii) free fermions in Hofstadter's butterfly.
}

\maketitle

\section{Introduction}
The behavior of a two-dimensional electron system under a magnetic field is a simple
 but at the same time particularly significant issue in the field of condensed matter physics. 
Previous studies have discovered dozens of important properties by considering non-interacting systems.
The Hofstadter problem, which deals with a tight-binding model on an isotropic square lattice under a uniform magnetic field,
 is one of the most intriguing problems\cite{hofstadter}.
While analytic solutions for a proper equation do not exist for a general magnetic field\cite{harper}, two extraordinary properties are known for this system: 
the integer quantum Hall effect (IQHE) at zero temperature and the fractal energy band structure, which is usually referred to as ``Hofstadter's butterfly''.

Since the discovery of the IQHE by von Klitzing {\it et al.}, the quantum Hall effect (QHE) has  continued to draw much interest\cite{klitzing,tsui_stormer_gossard,zhang_2005,ando_uemura,laughlin,tknn,kohmoto_top,zheng_ando,gusynin}. 
There is no doubt that this strikingly odd phenomenon has contributed to the development of condensed matter physics both theoretically and experimentally. 
For instance, the validity of gauge analysis has been recognized since Laughlin described the IQHE from the perspective of gauge invariance\cite{laughlin}. 
Thouless {\it et al.} (or TKNN) stated that a non-interacting two-dimensional electron system exhibits the IQHE
when the chemical potential is in the energy gap
 \cite{tknn}, which was later found to account for the gauge-invariant nature of $\sigma_{xy}$\cite{kohmoto_top}. 
Namely, the quantization of the Hall conductance  can be identified as a Chern number, a topological invariant quantity whose value is determined from the so-called ``Berry connection''\cite{berry,xiao}. Other calculation techniques are the Streda formula \cite{streda} and the adoption of lattice gauge theory \cite{fukui,hatsugai}, both of which have been found to be consistent so far.

It is implied from the robustness of $\sigma_{xy}$ unless the gaps open and close \cite{avron,oshikawa} that the label of each gap is related to the value of $\sigma_{xy}$. Surprisingly, this naive expectation is realized at least in a square lattice. Let the magnetic flux per plaquette $\phi$ be a rational number in units of flux quantum $\phi_0$, i.e.,  $\phi/\phi_0=p/q$ ($p$ and $q$ are coprime). 
Dana {\it et al.} \cite{dana}proposed that magnetic translational symmetry yields the simple relationship $p\sigma_{xy} +2qm=2r$
 ($m,r=$integers) when the Fermi energy lies in the $r$th gap from the bottom.  
Note that spin degeneracy is included in this so-called ``Diophantine equation''.
Among the multiple candidates of $(\sigma_{xy},m)$ that satisfy this equation, the solution can be uniquely determined by imposing the condition $|\sigma_{xy}|\leq q$.
Until now, no counterexample for this condition has been found. 
However, the physical justification is still under discussion. 
For instance, TKNN claimed that the discussion in the anisotropic case can be extended to the isotropic case, while they admit that there is no proof of this.\cite{tknn,kohmoto_diophantine} 
Chang and Niu considered the Einstein--Brillouin--Keller quantization, an extended method of Bohr--Sommerfeld quantization, of wave packets to obtain the Chern number of each bands.\cite{chang_niu_1995,chang_niu_1996} While the continued fraction expansion (CFE) of $\phi$ plays an important role in their theory, this expansion differs from the one used to describe the self-similarity of Hofstadter's butterfly. In other words, the relationship between the fractality and the quantization rule is unclear.

In this paper, we interpret the peculiar quantization rule in two different ways. For any rational $\phi$, both interpretations are valid.
One is based on the connection to the Landau quantization of massless Dirac fermions, 
which is realized at nearby $\phi' (\neq \phi) $ with even $q$. 
In this case, owing to the $q$-fold degeneracy of the Brilluoin zone, 
the Hall conductance is $\sigma_{xy}=\frac{e^2}{h}q(N+\frac{1}{2})\ $($N=$integer) 
near zero energy\cite{gusynin},
where $N$ denotes the filling factor. The quantization rule for a specific $\phi$ can be given by considering such particles originating from $\phi'$.
On the other hand, we can also connect the quantization rule with the Laudau quantization of free fermions, which is realized at $\phi''$ with odd $q$.
This phenomena in turn 
yields $\sigma_{xy}=\frac{e^2}{h}qN\  $
({$N=$}integer), with $N$ being the filling factor again.
 
Admitting that two interpretations deal with the energy gaps between subcells, we emphasize that the consistency with the Diophantine equation holds.
Later, we see in detail that such $q$'s relate the Diophantine equation and the fractality of Hofstadter's butterfly.

This paper is organized as follows. Section 2 gives brief information on Hofstadter's butterfly. In particular, the notion of the ``subcell'' is  essential in understanding the fractal structure of the diagram. In Sect. 3, we newly define ``family'' to give a clear statement on the relationship between the IQHE and the fractality in Sect. 4. Finally, the conclusion is given in Sect. 5.

\section{Fractal Structure of the Butterfly}
The Hamiltonian of the Hofstadter problem is
\begin{eqnarray}\label{hamiltonian}
H &=& -t\sum_{\langle i,j \rangle,\sigma} c_{i,\sigma}^{\dagger}c_{j,\sigma}e^{i\theta_{ij}}+h.c.,
\end{eqnarray}
where $c_{i,\sigma}^{\dagger}$ ($c_{i\sigma}$) is a creation (annihilation) operator of an electron of  spin $\sigma$ on the $i$ site, 
and $t$ denotes a nearest-neighbor transfer integral. 
$\theta_{ij}$ is Peierls phase and for each plaquette, 
$\underset{\text{plaquette}}{\Sigma}\theta_{ij} = 2 \pi \phi$ holds, 
where $2\pi\phi$ is the magnetic flux per plaquette. 
Taking $e=h=1$, we can rewrite the Hamiltonian for the rational flux $\phi=p/q$ ($p$ and $q$ are coprime) as
\begin{eqnarray}\label{Harper}
H &=& -t\sum_{{\bf k}\in \mathrm{BZ}}\widetilde{\mathbf{c}}^{\dagger}(\mathbf{k})\widetilde{H}(\mathbf{k})\widetilde{\mathbf{c}}(\mathbf{k}),
\end{eqnarray}
where $\widetilde{H}(\mathbf{k})$ is a $q \times q$ matrix whose matrix elements are given by 

\begin{widetext}
\begin{equation}\label{hamiltonian_k}
\widetilde{H}({\bf k}) = 
\left( 
 \begin{array}{ccccc}
 2\mathrm{cos}(k_x)&1&0&\cdots &e^{-iqk_y}\\
 1&2\mathrm{cos}(k_x + 2\pi \phi)&1&\cdots &0\\
 0&1&2\mathrm{cos}(k_x + 4\pi \phi)&\cdots &0\\
 \vdots&&&\ddots&\\
e^{iqk_y}&0&0&\cdots &2\mathrm{cos}(k_x + 2(q-1)\pi \phi)
 \end{array}
\right),
\end{equation}
\end{widetext}

with
\begin{equation}
\widetilde{\mathbf{c}}(\mathbf{k}) = {}^{t}\!{\left(\widetilde{c}_{0}(\mathbf{k}),\widetilde{c}_{1}(\mathbf{k}),...,\widetilde{c}_{q-1}(\mathbf{k})\right)}.
\end{equation}
\begin{figure}[hlb]
 \begin{center}
     \resizebox{7cm}{5cm}{\includegraphics{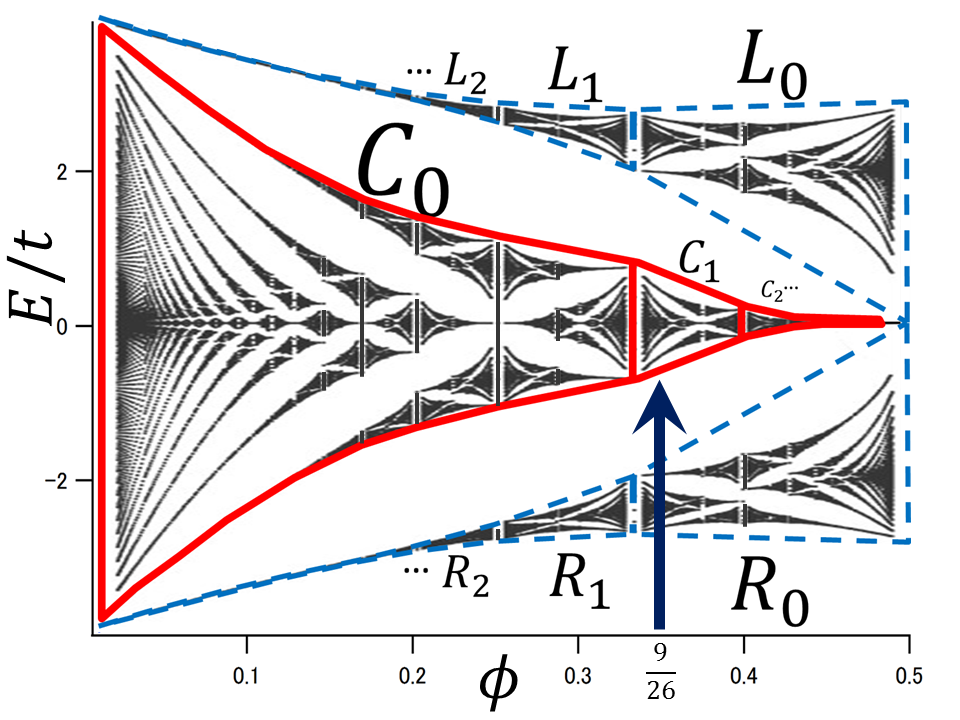}}\\ 
      \centering \ \ \ \ (a) \\
     \resizebox{7cm}{5cm}{\includegraphics{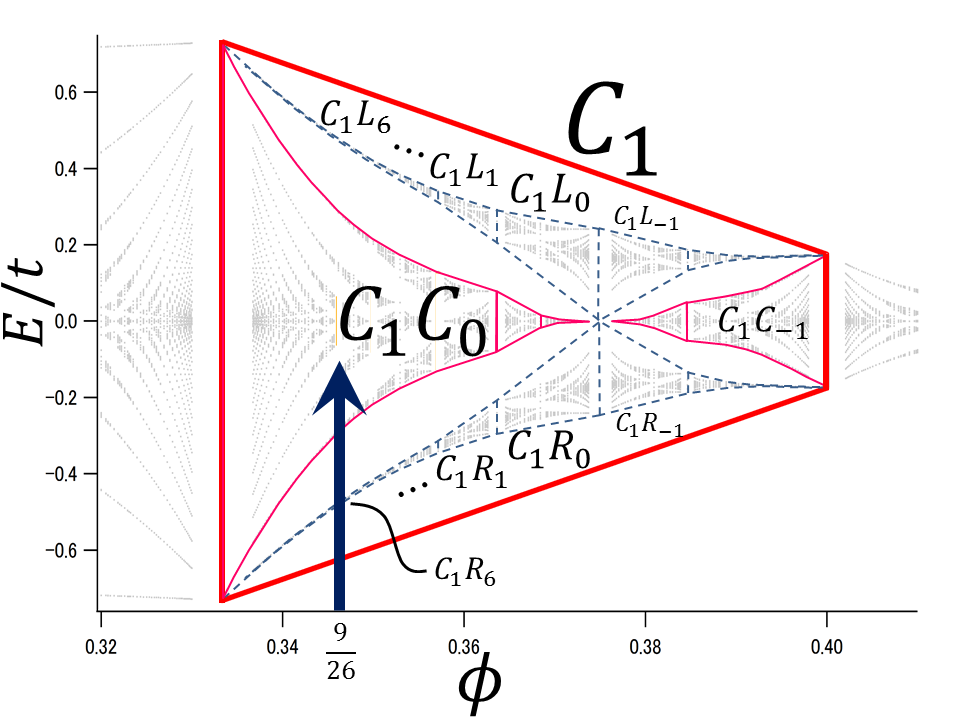}}\\
      \centering \ \ \ \ (b)
 \end{center}
     \caption{\label{coloredbutterfly}(Color online) (a) Left half $(0 \leq \phi \leq 1/2)$ of the butterfly divided into subcells. The solid and dashed lines indicate the boundary of $C$ subcells and that of $L, R$ subcells, respectively. (b) Expanded energy spectrum inside the $C_1$ subcell. The arrows indicate $\phi =9/26$, whose 26 energy bands are classified  in Sect.  \ref{familysection} into families.}
\end{figure}
$\widetilde{c_m}(\mathbf{k})$ ($0\leq m\leq q-1$) denotes the fermionic annihilation operator in the reciprocal space,
\begin{eqnarray}
 \widetilde{c}_{m}(\mathbf{k})=\frac{1}{\sqrt{L_x}}\frac{1}{\sqrt{L_y/q}}\sum_{n,m'}e^{-ik_xn-ik_yqm'}c_{n,qm'+m},
\end{eqnarray}
where the indices $n$ and $qm'+m$ represent the $x-$ and $y-$coordinates of the lattice point, respectively.
Equation (\ref{Harper}) is known as Harper's equation\cite{harper}, and each band of this Hamiltonian is $q$-fold degenerate along the $k_x$-axis.

Figure 1(a) shows the energy spectrum as a function of $\phi$ for $0 \leq \phi \leq 1/2$, which corresponds to the left half of Hofstadter's butterfly. 
Red and blue lines denote the boundary of the subcell, which was introduced by Hofstadter in order to describe the self-similar structure of the butterfly\cite{hofstadter}. 
A subcell is defined as a part of the graph whose structure is self-similar to the original one.
In the following, we review the properties of subcells to prepare for the argument in the following sections.
Our argument is restricted to $0\leq \phi \leq 1/2$ since the butterfly is symmetric with respect to $\phi$=1/2.

 There are three types of subcells: $L, R$, and $C$ [See Fig. \ref{coloredbutterfly}(a)]. $L$ and $R$ subcells are the outermost energy bands, which are surrounded by blue lines.
$L_n$ and $R_n$ ($n=0, 1, 2 \cdots$) are defined in the regions of 
\begin{equation}
\frac{1}{n+3}<\phi\leq\frac{1}{n+2}. 
\end{equation}
$C$ subcells are the center bands surrounded by red lines in Fig. \ref{coloredbutterfly}(a), and $C_n$ ($n=0, 1, 2 \cdots$) are in the regions of 
\begin{equation}
\frac{n}{2n+1}\leq\phi<\frac{n+1}{2n+3}.
\end{equation}
Note that the existence of energy gaps between $L$, $C$, and $R$ subcells enables us to  divide the diagram uniquely.

As claimed by Hofstadter\cite{hofstadter} and proved later by MacDonald\cite{macdonald}, an appropriate linear stretching of the variable 
\begin{equation}\label{betadef}
\beta = \frac{1}{\phi} -\left[\frac{1}{\phi}\right]
\end{equation}
 almost deforms $L_n$ and $R_n$ into the original butterfly. Here, the term ``almost'' reflects the slight difference; the minibands in $L_n$ and $R_n$ are gapped at $\beta = 1/2, 1/4$, etc., while the gaps in the original butterfly close at $\phi = 1/2, 1/4$, etc. However, this is a minor difference since our argument is restricted to the $C$ subcells in the following. $\beta$ is called the ``local variable''. For the case of $C_n$, a similar local variable,
\begin{equation}\label{betadef2}
\beta=\frac{1}{\phi^{-1}-2}-\left[\frac{1}{\phi^{-1}-2}\right],
\end{equation}
deforms each subcell into the original butterfly.

Next, to discuss the detailed structure of the butterfly, 
we denote the subcells inside $C_n$ as $\widetilde{L}_m, \widetilde{R}_m$, and $\widetilde{C}_m$. 
Then, we denote  ``subcell $\widetilde{C}_m$ inside subcell $C_n$''  as $C_nC_m$ in the following.
For instance, Fig. 1(b) shows the subcells inside $C_1$.
In this way, we have a simple rule to divide the butterfly spectrum into smaller areas. In the following, we continue to divide $C$ subcells only. 
The repetition of this operation determines the division of the butterfly spectrum uniquely. 
In the next section, we define ``family'' based on the above argument.

 \section{Grouping Rule for ``Families''} \label{familysection}

\begin{figure}[t]
 \begin{center}\centering
    \resizebox{\columnwidth}{!}{\includegraphics{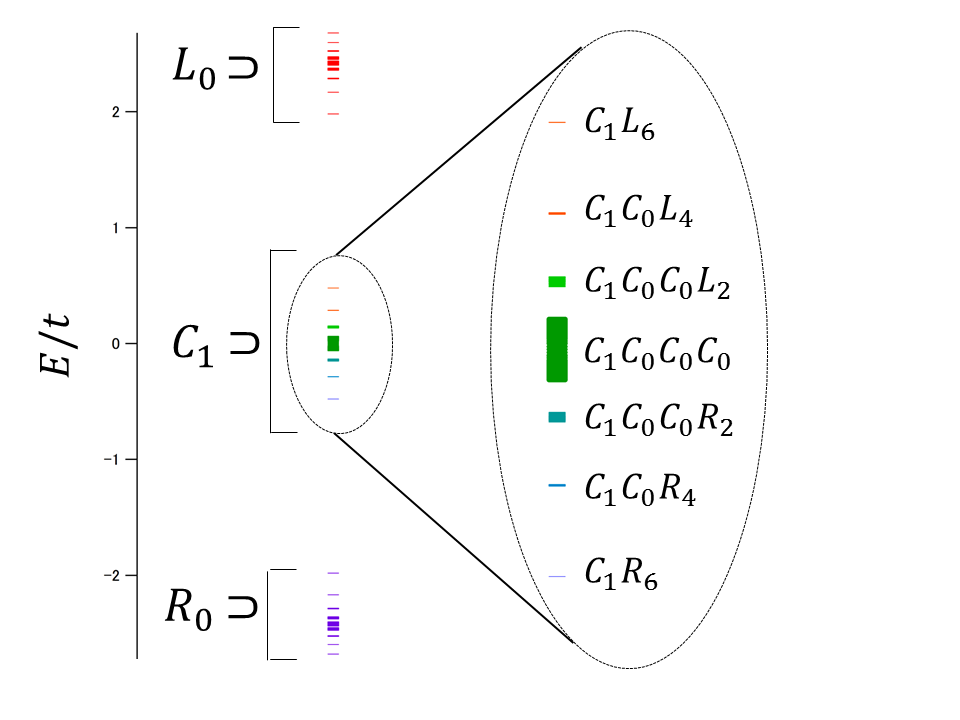}}
    \caption{\label{family_label}Energy spectrum at $\phi=\frac{9}{26}$. Twenty-six bands are divided into families.}
 \end{center}
\end{figure}

 Next, we introduce the concept of ``family'' in the butterfly. 
 By using families, we can understand the relationship between the fractality and the quantization rule of $\sigma_{xy}$.

We define the set of energy bands in a subcell $X$ as a family $X$ for a fixed $\phi$. 
For explanation, we use an example at $\phi=9/26$, which has 26 bands. 
Figure \ref{family_label} shows the energy spectrum obtained at $\phi=9/26$. 
Since $\phi=9/26$ is slightly larger than 1/3, the outermost bands belong to $L_0$ and $R_0$, as shown in Fig. \ref{coloredbutterfly}(a) by an arrow, while the other bands are included in $C_1$. 
Therefore, families $L_0$ and $R_0$ contain 9 bands while family $C_1$ contains 8 bands.
The central region of the energy spectrum (family $C_1$) is expanded in the r.h.s. of Fig. \ref{family_label}. 
By comparing with Fig. \ref{coloredbutterfly}(b), we can see that the outermost bands in $C_1$ correspond to $\widetilde{L}_6$ or $\widetilde{R}_6$ inside $C_1$, which are denoted as families $C_1L_6$ and $C_1R_6$, respectively. (In this case, the family has only a single band.) 
In the same way, we consider other bands near $E=0$. 
From Fig. \ref{coloredbutterfly}(b), we can see that these bands belong to $\widetilde{C}_0$ inside $C_1$, i.e., $C_1C_0$. 
Therefore, this $C_1C_0$ family has 6 bands.
Furthermore, the outer two bands in $C_1C_0$ are identified as $\widetilde{L}_4$ or $\widetilde{R}_4$ in $C_1C_0$, so that they are expressed as $C_1C_0L_4$ and $C_1C_0R_4$ and the corresponding family contains a single band.  
In a similar way, we can identify families $C_1C_0C_0L_2, C_1C_0C_0C_0$, and $C_1C_0C_0R_2$.
Note that family $C_1 C_0 C_0 C_0$ has two bands that touch at $E=0$.

Next, we use $\Gamma_m(\beta)$ and $\Lambda_l(\beta)$ ($m,l=$ integer), which were introduced by Hofstadter to describe the fractality of $C$ subcells and $L(R)$ subcells\cite{hofstadter}. We will show that $\Gamma_m (\beta)$ and $\Lambda_l (\beta)$ give the number of bands in each family.
The definitions for $m,l \geq 0$
are
\begin{eqnarray}
\Gamma_{m}(\beta) &=& \frac{1}{2+(m+\beta)^{-1}},\ 0 \leq \beta <1\label{leftcfe_c}\\
\Lambda_{l}(\beta) &=& \frac{1}{(l+2)+\beta}.\label{leftcfe_lr}
\end{eqnarray}

 For example, $\phi=9/26$ can be expressed as various CFEs as follows:

\begin{eqnarray}\label{CFEof9_26}
\displaystyle
\frac{9}{26} = 
\begin{cases}
\frac{1}{(2+0) + \frac{8}{9}} = \Lambda_0 \left(\frac{8}{9}\right)\\
\frac{1}{2 + (1+\frac{1}{8})^{-1}} = \Gamma_1 \left(\frac{1}{8}\right) =
\begin{cases} 
\frac{1}{2+(1+\frac{1}{(2+6)+0})^{-1}} = \Gamma_1\Lambda_6(0)\\ 
\frac{1}{2+(1+\frac{1}{2+(0+\frac{1}{6})^{-1}})^{-1}} = \Gamma_{1,0} \left(\frac{1}{6}\right)
\end{cases}
\end{cases}\nonumber
\end{eqnarray}
\begin{eqnarray}\hspace{-5mm}
\Gamma_{1,0}\left(\frac{1}{6}\right) = 
\begin{cases}
\Gamma_{1,0}\Lambda_4(0)\\
\Gamma_{1,0,0}\left(\frac{1}{4}\right) =
\begin{cases}
\Gamma_{1,0,0}\Lambda_2(0)\\
\Gamma_{1,0,0,0}\left(\frac{1}{2}\right).
\end{cases}
\end{cases}
\end{eqnarray}
Here, $\displaystyle\Gamma_{m_1,m_2\cdots m_N}(\beta)$ and $\Gamma_{m_1,m_2\cdots m_N}\Lambda_l(\beta)$ represent the continued fractions $\Gamma_{m_1}\{\Gamma_{m_2}\{
\cdots[\Gamma_{m_N}(\beta)]\cdots\}\}$ and $\Gamma_{m_1}\{\Gamma_{m_2}
\{\cdots\{\Gamma_{m_N}[\Lambda_l(\beta)]\}\cdots\}\}$, respectively. 
We emphasize that the role of the CFE is to subdivide families in detail.
We find that the indices and arguments of $\Gamma$ and $\Lambda$ correspond to the name of the family and the number of bands in the family, respectively. 
For instance, $\phi = \Lambda_0 \left(\frac{8}{9}\right)$ tells us that 9 bands belong to both families $L_0$ and $R_0$. 
This is because the transformation in Eq.(\ref{betadef}) gives
\begin{equation}
\beta = \frac{1}{\phi}-\left[\frac{1}{\phi}\right] =\frac{26}{9}-2=\frac{8}{9}\ \ , \mathrm{i.e.},\ \ \phi=(2+\beta)^{-1},
\end{equation}
and $\beta$ represents the effective flux in $L_0$, whose denominator represents the number of bands in family $L_0$.
Similarly, $\phi=\Gamma_1\left(\frac{1}{8}\right)$ implies that the remaining 8 bands belong to family $C_1$. 
This is because the transformation in Eq. (\ref{betadef2}) gives
\begin{equation}
\beta = \frac{1}{\phi^{-1}-2}-\left[ \frac{1}{\phi^{-1}-2}\right]=\frac{9}{8}-1=\frac{1}{8},
\end{equation}
and $\beta$ represents the effective flux in $C_1$. In a similar way, we can determine the number of bands belonging to each family. For example, $\Gamma_1\Lambda_6(0)$ means that this family has only one band since we interpret $\beta=0$ as $p=0$ and $q=1$. In other words,  $\beta =0$ means that the effective flux is zero so that there is a single energy band.

In this way, we can continue the CFE until the effective flux $\beta$ becomes 0 or 1/2. If $\beta=0$, the corresponding family has a single band, while if $\beta$=1/2, the corresponding family has two bands. This CFE should end at some state since the number of bands is finite for a rational flux $\phi=p/q$. 

Equation (\ref{CFEof9_26}) shows the detailed CFE for $C_1$. Here, let us mention the CFE for $L_0$ or $R_0$, i.e., $\Lambda_0 \left(\frac{8}{9}\right)$. 
The local variable is {$\beta=8/9>1/2$}, at which Eqs.
(\ref{leftcfe_c}) and ({\ref{leftcfe_lr}}) cannot be used. Therefore, for
$m,l \leq 0$,
we introduce

\begin{eqnarray}
\Gamma_{-m'-1}(\beta) &=& 1- \frac{1}{2+(m'+\beta)^{-1}}, \ \ m'\geq0\\
\Lambda_{-l'-1}(\beta) &=& 1- \frac{1}{(l'+2)+\beta}, \ \ l'\geq0,
\end{eqnarray}
which should be used when the effective flux is larger than 1/2. Using these notations, we obtain
\begin{eqnarray}
\displaystyle
\Lambda_0\left(\frac{8}{9}\right) =
\begin{cases}
\Lambda_0 \Lambda_{-8}(0)\\
\Lambda_0 \Gamma_{-1}\left(\frac{1}{7}\right)=
\begin{cases} 
\Lambda_0 \Gamma_{-1} \Lambda_5(0)\\ 
\Lambda_0 \Gamma_{-1,0} \left(\frac{1}{5}\right) 
\end{cases}
\end{cases}\nonumber
\end{eqnarray}

\begin{eqnarray}\hspace{0.2cm}
\Lambda_0 \Gamma_{-1,0} \left(\frac{1}{5}\right) =
\begin{cases}
\Lambda_0 \Gamma_{-1,0}\Lambda_3(0)\\
\Lambda_0 \Gamma_{-1,0,0}(\frac{1}{3})=
\begin{cases}
\Lambda_0 \Gamma_{-1,0,0}\Lambda_1(0)\\
\Lambda_0 \Gamma_{-1,0,0,1}(0).
\end{cases}
\end{cases}
\end{eqnarray}
This means that family $L_0$, which contains 9 bands,
 can be decomposed into 9 families, each of them having only one band.
For instance, $\Lambda_{0}\Lambda_{-8}(0)$ corresponds to families $L_0L_{-8}$ and $L_0R_{-8}$, each of which contains a single band. 
Likewise, {$\Lambda_0 \Gamma_{-1}\Lambda_5(0)$}, etc., correspond to two families {$L_0 C_{-1}L_5$} and {$L_0 C_{-1}R_5$}, etc., and {$\Lambda_0\Gamma_{-1,0,0,1}(0)$} corresponds to  family {$L_0 C_{-1}C_0C_0C_1$}. All of them contain only one band as well.


\section{Relationship between Hall Conductance and Energy Spectrum}
\subsection{Massless Dirac fermions}
\begin{figure}[b]
 \begin{center}\centering
  \resizebox{\columnwidth}{!}{\includegraphics{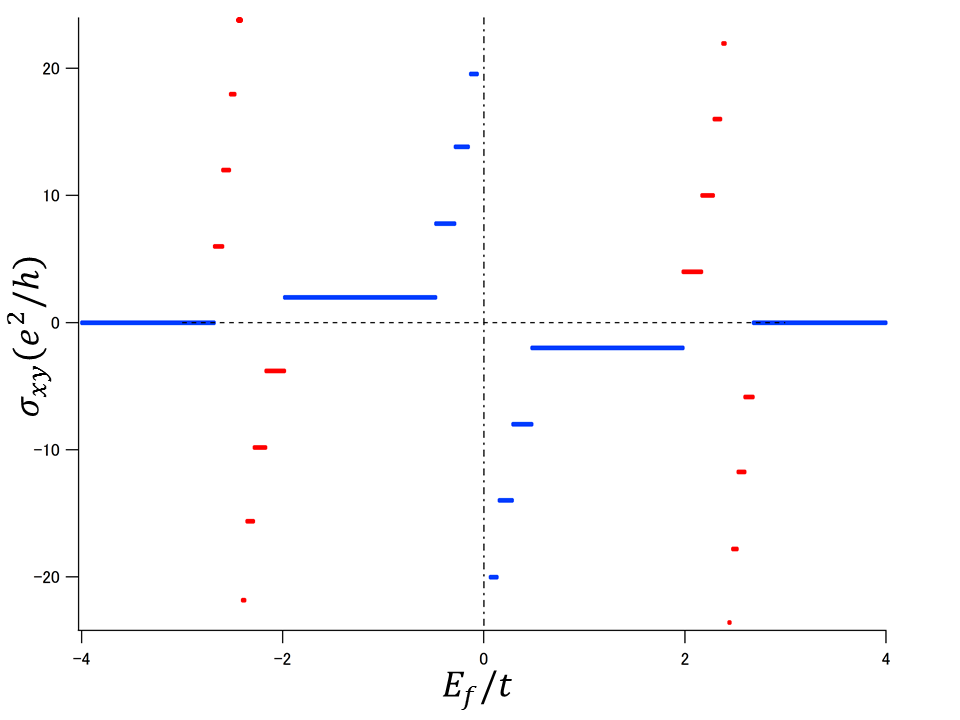}}
  \caption{Hall conductance as a function of the Fermi energy for $\phi = 9/26$. }
  \label{sigma_of_family2}
 \end{center}
\end{figure}

In this section, we discuss the relationship between $\sigma_{xy}$ (or the Chern number) and families. We numerically calculate the Hall conductance at $\phi=9/26$, which is shown in Fig. \ref{sigma_of_family2}. (Here, we have included the spin degeneracy of 2 in the calculation of $\sigma_{xy}$.)
We find a definite rule between the absolute value of $\sigma_{xy}$ and the CFE obtained in the previous section.

Let us consider the family just below the energy gap when the Fermi energy is located in a gap. 
Assume that the corresponding family has a CFE of $\Gamma_{m_1\cdots}(\beta)$. 
Then, we put $\beta=1/2$ artificially. We find that the denominator of $\Gamma_{m_1\cdots}\left(\frac{1}{2}\right)$ is equal to the absolute value of $\sigma_{xy}$. 
For example, we obtain $\Gamma_1\left(\frac{1}{2}\right)$ = 3/8, $\Gamma_{1,0}\left(\frac{1}{2}\right) = 5/14$, and $\Gamma_{1,0,0}\left(\frac{1}{2}\right) = 7/20$ from Eq. (\ref{CFEof9_26}) when the Fermi energy is located just above families $C_1L_6, C_1C_0L_4, C_1C_0C_0L_2,$ respectively. 
The denominators, 8, 14, and 20 of $\Gamma_1\left(\frac{1}{2}\right)$, $\Gamma_{1,0}\left(\frac{1}{2}\right)$, and $\Gamma_{1,0,0}\left(\frac{1}{2}\right)$, correspond to the absolute values of $\sigma_{xy}$ in the region of $0<E_f/t<0.5$ in Fig. \ref{sigma_of_family2}.

We give a physical interpretation of this peculiar rule by discussing the Landau quantization of massless Dirac fermions. 
In the following, we refer to the Landau levels (LLs) of such particles as ``Dirac LLs''\cite{hatsugai}. 
Let us first consider the case with $\sigma_{xy}=-2$, which is realized when $E_f$ is located between the families $C_1$ and $L_0$, i.e., $0.5<E_f/t<2$ in Fig. \ref{sigma_of_family2}. 
From the global structure shown in Fig. \ref{coloredbutterfly}(a), we can see that this gap continues to $\phi=1/2$. 
At $\phi=1/2$, the eigenstate is the so-called ``$\pi$-flux state'', at which two bands cross linearly, i.e., form a Dirac cone, at zero energy. 
As we show in the Appendix, this model can be effectively described by four (2+1)-dimensional massless Dirac fermions in the vicinity of $E\sim0$. 
When a small magnetic flux is imposed (i.e., $\phi=1/2 +\Delta \phi $),  $\sigma_{xy}$ is quantized as 
\begin{eqnarray}\label{graphene}
\sigma_{xy}=sgn(\Delta\phi)\cdot4\cdot(n+1/2),
\end{eqnarray} 
which has an identical expression to the anomalous Hall effect in the honeycomb lattice \cite{gusynin}. 
Namely, the model is equivalent to a four (2+1)-dimensional massless Dirac fermion system under an effective magnetic flux $\Delta \phi$. 
Actually, the energy dispersion near $\phi=1/2$ in Fig. \ref{coloredbutterfly}(a) shows the typical behavior $E_n\propto \pm \sqrt{n}$ for the Dirac LLs\cite{mcclure}. 
Although the $n=0$ Dirac LL broadens as $\phi$ decreases from $\phi=1/2,$ it remains detached from other LLs and forms $C_1$.
Since the Chern number is robust unless the band gaps open and close \cite{oshikawa}, $\sigma_{xy}$ of LLs is conserved, which explains why $\sigma_{xy}=-2$ always holds when $E_f$ is between $C_1$ and $L_0$.

$\sigma_{xy}=-8$ can be understood in a similar way; the self-similarity of the diagram and the CFE, $9/26 = \Gamma_1\left(\frac{1}{8}\right)$ in Eq. (\ref{CFEof9_26}), indicate that the effective flux is equal to 1/8 in $C_1$ [see Fig. \ref{coloredbutterfly}(b)].
Then, from Fig. \ref{coloredbutterfly}(b), we can see that family $C_1C_0$ corresponds to the $n=0$ Dirac LL connected to the state at $\phi=3/8$, which can be obtained by setting the local variable $\beta$ as 1/2, i.e., $\Gamma_1(\beta=1/2)=3/8$. 
Since the denominator is 8, this means that the effective model is described by 16 massless Dirac fermions. 
Since  $C_1C_0$ corresponds to the $n=0$ Dirac LL, the Hall conductance is -8 when $E_f$ is located in the gap just above it. Furthermore, we obtain $\sigma_{xy}=-14$ and -20 by considering the quantization of $\sigma_{xy}$ by 28 and 40 Dirac fermions, respectively.

For general cases with $\phi = \Gamma_{m_1\cdots m_N}(\beta^{(N+1)})$, we obtain a series of integers $\{ Q_1,...,Q_N \}$ by calculating $1/2=P_1/Q_1, \Gamma_{m_1}\left(\frac{1}{2}\right)=P_2/Q_2, \cdots ,$ and $\Gamma_{m_1\cdots m_{N-1}}\left(\frac{1}{2}\right)=P_N/Q_N$. 
Similar to the discussion for $\phi=9/26$, the gap between the $(l+1)$th and $l$th families from the top corresponds to the gap above the $n=0$ Dirac LL generated from $\phi=P_l/Q_l$. 
Therefore, the Hall conductance of the $l$th family from the top $\sigma_{xy}(l)$ is given as

\begin{eqnarray}\label{gapcond}
\sigma_{xy}(l) = sgn\left(\phi - \frac{P_l}{Q_l}\right)\cdot2Q_l\cdot \frac{1}{2}, \nonumber
\end{eqnarray} 
where the factor $2Q_l$ is from the number of Dirac fermions, and $1/2$ comes from the (2+1)-dimensional chiral anomaly in $\sigma_{xy}$. 

\subsection{Free fermions}
Here, we give another interpretation for the quantization rule of $\sigma_{xy}$. In Sect. 4.1, we considered the ``$\pi$-flux state'' at $\beta=1/2$.
Here, we are interested in the opposite side, i.e., zero flux at $\beta=0$. 
We consider the total Chern numbers of all the energy bands in a specific family, which can be obtained by calculating $\sigma_{xy}(l)-\sigma_{xy}(l+1)$.
We show that this quantity is related to the LLs of nonrelativistic free fermions (or Fermi LLs in Hatsugai {\it et al.}'s notation\cite{hatsugai}). 

Let us use $\phi=9/26$ again. From Figs. \ref{coloredbutterfly} and \ref{sigma_of_family2}, we can see that the total Chern number for $L_0$ is equal to 2.
Similarly, we obtain 6 for families $C_1L_6, C_1C_0L_4,$ and $C_1C_0C_0L_2$. 
Let us start with the $L_0$ family [$\sigma_{xy}(1)- \sigma_{xy}(2)= 2$].
Again, from the global structure shown in Fig. \ref{coloredbutterfly}(a), we can see that $L_0$ is connected to the lowest Fermi LL at $\phi=0$.
Under a uniform magnetic flux $\Delta\phi$, the Hall conductance is quantized as 
\begin{eqnarray}
\sigma_{xy}=sgn(\Delta\phi)\cdot2\cdot n.
\end{eqnarray}
$\sigma_{xy}$ of each LL is 2 including the spin degeneracy, which explains why the Chern number of family $L_0$ is equal to 2.

A similar discussion holds for the other families. 
The self-similarity of the diagram and the CFE, $9/26=\Gamma_1\left(\frac{1}{8}\right)$, means that the effective flux is 1/8 in $C_1$ as discussed previously.
If we set the local variable $\beta$ as $\beta=0$, then $\Gamma_1(\beta=0)=1/3$ means that families $C_1L_6, C_1C_0L_4$, and $C_1C_0C_0L_2$ all correspond to the Fermi LLs of free fermions at $\phi=1/3.$ 
The denominator 3 of $\phi$ means that the effective model is a three-nonrelativistic-free-fermion system. 
Hence, the Chern number of families $C_1L_6$, etc., is 6.

For general cases with $\phi$, we consider a series of integers $\{Q'_1,...,Q'_N\}$ by computing $0/1=P'_1/Q'_1,\Gamma_{m_1}(0)=P'_2/Q'_2, \cdots , \Gamma_{m_1\cdots m_{N-1}}(0)=P'_{N}/Q'_N$. 
The Chern number of the $l$th family from the top satisfies
\begin{eqnarray}\label{familycond}
\sigma_{xy}(l)-\sigma_{xy}(l+1)=sgn\left(\phi-\frac{P'_l}{Q'_l}\right)\cdot2Q'_l\cdot1,
\end{eqnarray}
where the factor $2Q'_l$ stands for the number of free fermions including spin degeneracy, and 1 corresponds to the contribution of each Fermi LL.

\section{Conclusions}
We have introduced a new notion, ``families'', to classify the bands at specific $\phi$ in the Hofstadter butterfly, which provides two types of physical interpretation for the quantization rule of $\sigma_{xy}$. 
One is based on the Landau quantization of (2+1)-dimensional massless Dirac fermions, and the other of nonrelativistic free fermions. 
In the former approach, we showed that the Landau quantization of massless Dirac fermions explains the $\sigma_{xy}$ in family gaps. 
On the other hand, the Chern number  of each family can be obtained by considering the ordinary Landau levels of free fermions in the latter case.

\section*{Acknowledgements}
The authors wish to thank Y. Hatsugai and H. Katsura for many fruitful discussions. This work was supported by a Grant-in-Aid for Scientific Research on
\lq\lq Multiferroics in Dirac electron materials'' (No.\ 15H02108).

\section*{Appendix}
There is a unitary matrix $S$ that anticommutes with the hopping Hamiltonian when the lattice is bipartite, 
\begin{eqnarray}\label{property_of_S}
S^{\dag}HS = -H.
\end{eqnarray} 
Note that $S$ transforms an energy eigenstate into an opposite-signed energy eigenstate. As $E=0$ is unchanged by $E \rightarrow -E$, we can choose the energy eigenstates of $E=0$, $\ket{\phi_i}$ ($i=1,\cdots, q_0$), to be the eigenstates of $S$. 
In other words, $ \ket{\phi_i}$ are simultaneously eigenstates of $H$ and $S$. Owing to Eq. (\ref{property_of_S}), we can calculate the diagonal term of $H$ as 
\begin{eqnarray}\label{antiS}
\Braket{\phi_i|H|\phi_i}=\left(\bra{ \phi_i} S^{\dag}\right) H \Big( S\ket{\phi_i} \Big)\nonumber\\
=\Braket{\phi_i|S^{\dag}HS | \phi_i}=\Braket{\phi_i|-H|\phi_i}=0.
\end{eqnarray}

We will show in the following argument that, by using Eq. (\ref{antiS}), the Hofstadter problem in the square lattice with $\phi=p/q$ with an even $q$ reduces to $2q$ massless Dirac fermions in the continuum limit. 
As Kohmoto has shown\cite{kohmoto}, there are $q$ Dirac points, ${\bf k}={\bf K}_j = (0,2\pi j/q)$ $(j=1,\cdots,q)$, at which two bands, $\ket{\psi_1(\bf k)}$ and $\ket{\psi_2(\bf k)}$, cross each other. 
Since the square lattice is bipartite, we can choose the eigenstate at ${\bf K}_j$, $\ket{\psi_i({\bf K}_j)}$ ($\widetilde{H}({\bf K}_j)\ket{\psi_i({\bf K}_j)}=0$) to be the simultaneous eigenstates of $q\times q$ matrices $\widetilde{H}({\bf K}_j)$ and $S$.
Then, by substituting $H=\widetilde{H}({\bf k})$ and $\ket{\phi_i}=\ket{\psi_i({\bf K}_i)}$ into Eq. (\ref{antiS}), we can immediately conclude that the expectation values for $\widetilde{H}({\bf k})$ are zero, 
\begin{eqnarray}\label{diagonalK}
\Braket{\psi_i({\bf K}_j)|\widetilde{H}({\bf k})|\psi_i({\bf K}_j)}=0.
\end{eqnarray}

Then, we expand the Hamiltonian around ${\bf K}_j$ up to the linear order, 
\begin{eqnarray}\label{expandham}
\widetilde{H}({\bf k}) &\sim&  \widetilde{H}({\bf K}_j) + \delta H(\delta{\bf k}),\nonumber \\
\delta H(\delta{\bf k}) &=& \delta k_x M + i\delta k_y A,
\end{eqnarray}
where ${\bf k} = {\bf K}_j+\delta{\bf k}$. 
The $k_x$ dependence of the Hamiltonian, which is expressed in Eq. (\ref{hamiltonian_k}), can be seen in diagnonal terms, so $M$ is real-diagonal. 
On the other hand, the $(1,q)$ and $(q,1)$ elements in Eq. (\ref{hamiltonian_k}), i.e., $e^{\pm iqk_y}$, are the only $k_y$-dependent matrix elements.  
These terms become real numbers at Dirac points, which satisfy $k_y=K_{j,y}=2\pi j/q$\cite{kohmoto}. Hence, $A$ is real and anti-symmetric.

Next, we consider the effective 2$\times$2 Hamiltonian, $\widetilde{H}_{\mathrm{eff}}(\delta{\bf k})$,  in the restricted Hilbert space spanned by $\{\ket{\psi_1({\bf K}_j)},\ket{\psi_2({\bf K}_j)}\}$. 
We simply denote the eigenstates as $\ket{\psi_i^{(j)}}$ in the following.
By substituting Eq. (\ref{expandham}) into Eq. (\ref{diagonalK}), we obtain 
\begin{eqnarray}
\Braket{\psi_i^{(j)}|M|\psi_i^{(j)}}=\Braket{\psi_i^{(j)}|A|\psi_i^{(j)}}=0,
\end{eqnarray}
which causes the diagonal terms of the effective matrix to be zero.
The off-diagonal terms, $\Braket{\psi_{1(2)}^{(j)}|X|\psi_{2(1)}^{(j)}}$ ($X=M,A$), are non zero real numbers since we can choose $\ket{\psi_i^{(j)}}$ to be $q$-dimensional real vectors [$\widetilde{H}({\bf k}={\bf K}_j)$ is a $q\times q$ real matrix]. 
Therefore, the effective Hamiltonian is expressed as
\begin{eqnarray}
\widetilde{H}_{\mathrm{eff}}(\delta{\bf k})
=\delta k_x \sigma_x \Braket{\psi_1^{(j)}|M|\psi_2^{(j)}} + \delta k_y \sigma_y \Braket{\psi_1^{(j)}|A|\psi_2^{(j)}},
\end{eqnarray}
which reduces to the (2+1)-dimensional Dirac equation for massless fermions in the continuum limit. 

Finally, including the spin degeneracy, there are $2q$ Dirac points, so the effective Hamiltonian in the continuum limit can be described as 
\begin{eqnarray}
\widetilde{\mathscr{H}}_{\mathrm{eff}} = \sum_{a=1}^{2q}\overline{\psi}_a\partial_{\mu}\gamma_{\mu}\psi_a \ \ \ \ \ \ \ \ \ \ \ \  (\mu=1,2),
\end{eqnarray}
where $\gamma_{\mu}$ denotes the Pauli matrices $\sigma_x$ and $\sigma_y$.

\end{document}